\documentclass[twocolumn]{aastex63}

\usepackage{etoolbox}
\usepackage{graphicx}	% Including figure files
\usepackage{multirow}
\usepackage{cleveref}
\usepackage{CJK}
\usepackage{soul}

\newcommand{\kms}{km s$^{-1}$}

\newcommand{\lam}{$\lambda$}

\newcommand{\lya}{\mbox{Ly$\alpha$}}

\newcommand{\civ}{\mbox{C\,{\sc iv}}}
\newcommand{\ciii}{\mbox{C\,{\sc iii}}}
\newcommand{\cii}{\mbox{C\,{\sc ii}}}

\newcommand{\siii}{\mbox{Si\,{\sc ii}}}
\newcommand{\nv}{\mbox{N\,{\sc v}}}

\newcommand{\niii}{\mbox{N\,{\sc iii}}}

\newcommand{\ovi}{\mbox{O\,{\sc vi}}}

\newcommand{\mgi}{\mbox{Mg\,{\sc i}}}
\newcommand{\mgii}{\mbox{Mg\,{\sc ii}}}

\received{}
\revised{}
\accepted{}
\submitjournal{ApJ}

\shorttitle{Extreme High-velocity outflows}
\shortauthors{Chen et al.}
\graphicspath{{./}{figures/}}
\begin{document}

%\title{Absorption-Line Environments of High-Redshift BOSS Quasars}
\title{Extreme High-velocity outflows from High-Redshift BOSS Quasars}

\correspondingauthor{Bo Ma}
\email{mabo8@mail.sysu.edu.cn}

\author{Chen Chen}
\affiliation{School of Physics $\&$ Astronomy\\
Sun Yat-Sen University, Zhuhai Campus, \\ 
Zhuhai 519000, China}
\affiliation{Department of Physics $\&$ Astronomy\\
University of California\\ 
Riverside, CA 92521, USA}

\author{Fred Hamann}
\affiliation{Department of Physics $\&$ Astronomy\\
University of California\\ 
Riverside, CA 92521, USA}

\author{Bo Ma}
\affiliation{School of Physics $\&$ Astronomy\\
Sun Yat-Sen University, Zhuhai Campus, \\ 
Zhuhai 519000, China}

%\author{Britt Lundgren}
%\affiliation{Department of Astronomy,\\
%University of Wisconsin\\
%Madison, WI 53706, USA}

%\author{Donald York}
%\affiliation{Department of Astronomy $\&$ Astrophysics,\\
%University of Chicago\\
%Chicago, IL 60637, USA}

%\author{Daniel Nestor}
%\affiliation{Department of Physics $\&$ Astronomy,\\
%University of California\\
%Los Angeles, CA 90095, USA}

%\author{Yusra AlSayyad}
%\affiliation{Department of Astrophysical Sciences,\\
%Princeton University\\
%Princeton, NJ 08544, USA}



\begin{abstract}

It is common to assume that all narrow absorption lines (NALs) at extreme high-velocity shifts form in cosmologically-intervening gas or galaxies unrelated to the quasars. However, previous detailed studies of individual quasars have shown that some NALs at these large velocity shifts do form in high-speed quasar ejecta. We search for extreme high-velocity NAL outflows (with speeds $\sim0.1c-0.2c$) based on relationships to associated absorption lines (AALs) and broad absorption line (BAL) outflows. We find high-velocity NALs are strongly correlated with AALs, BALs and radio loudness, indicating that a significant fraction of high-velocity systems are either ejected from the quasars or form in material swept up by the radio jets (and are *not* unrelated intervening gas).
We also consider line-locked \civ\ doublets as another indicator of high-velocity NALs formed in outflows. The facts that line-locked NALs are highly ionized and correlated with BAL outflows and radio-loud quasars imply physical line-locking due to radiative forces is both common and real, which provide an indirect evidence that a significant fraction of high-velocity NALs are intrinsic to the quasars. 
%For example, previous studies of correlations between the incidence of high-velocity NALs and the quasar radio properties suggest that up to 36\% of these systems form in outflows.
%We search for extreme high-velocity NAL outflows (with speeds $0.1c-0.2c$) {\bf based on relationships to associated absorption lines (AALs) and broad absorption line (BAL) outflows, or based on large velocity offsets to AALs?}. We find High-velocity NALs are strongly correlated with AALs, BALs and radio loudness, indicating that a significant fraction of these systems are either ejected from the quasars or form in material swept up by the radio jets (and are *not* unrelated intervening gas). 
%which could be due to the lower densities in the infall absorbers to produce higher degrees of ionization in spite of the larger distances compared to outflow AALs.

\end{abstract}



\section{Introduction}

During the early stages of galaxy evolution, rapid star formation can be triggered by recent mergers of gas-rich galaxies, which is a key driving force of galaxy evolution over the history of the universe \citep{Sanders88, Elvis06, Hopkins08, Veilleux09}. In the centers of high-redshift massive galaxies, super-massive black holes (SMBHs) with high-Eddington ratio accretion can be identified as quasars. Powerful outflows driven by the quasar and/or the starburst can regulate the SMBH’s growth by cutting off the fuel supply and quench the star formation by expelling gas and dust from the galaxies \citep{Silk98, Kauffmann00, King03, Scannapieco04, DiMatteo05, Hopkins08, Ostriker10, Debuhr12, Rupke13, Rupke17, Cicone14, Weinberger17}.

%High-redshift quasars represent the active stages of black hole accretion and the evolution of massive galaxies in the early universe. They are very important to the study of the evolution of super-massive black holes (SMBHs) and their interaction with host galaxies. Some galaxy evolution models show that high-redshift quasars usually appear in the early stage of the evolution of massive galaxies and are formed after the merger of gas rich galaxies \citep{Sanders88, Elvis06, Hopkins08, Veilleux09}. Galaxy merging can trigger the growth of central black holes and the rapid star-formation in the host galaxies. The central black hole accretion and/or star formation will drive a powerful outflow, which can disperse gas and dust, so as to slow down the accretion and terminate star formation, which is the feedback mechanism of quasars\citep{Silk98, Kauffmann00, King03, Scannapieco04, DiMatteo05, Hopkins08, Ostriker10, Debuhr12, Rupke13, Rupke17, Cicone14, Weinberger17}. 

%The mechanism for high-speed quasar-driven outflows quenching star formation in their host galaxies is expected to involve the shredding and dispersal of interstellar clouds to produce complex outflowing gas structures on large scales \citep{Hopkins10, Faucher12b}. Infalling gas from the intergalactic medium (IGM) (e.g. cold mode accretion) is also believed to be important for fueling the central SMBH and star formation during these early active evolution stages \citep{Katz03, Keres09, Keres12}. 

The main tools used to study the gaseous environments of high-redshift quasars and test the models of massive galaxy evolution are absorption lines in quasar spectra. Quasar absorption lines give unique insight on the properties of the material associated with the quasar host galaxy or intervening galaxies along the line of sight, and are classified into broad absorption lines (BALs), narrow absorption lines (NALs) and mini-broad absorption lines (mini-BALs). For example, low-speed outflows or infall in the extended host galaxies should produce narrow associated absorption lines (AALs) with redshifts near the quasar emission-line redshifts and velocity widths less than a few hundred \kms \citep[e.g.,][]{Weymann79, Foltz86, Hamann97d, Hamann04, Simon10, Muzahid13, Chen20}. Cosmologically-intervening gas or galaxies unrelated to the quasars are detected via narrow absorption lines (NALs) at extreme velocity shifts ($\sim0.1c-0.2c$). High-speed quasar-driven outflows are detected via blueshifted broad absorption lines (BALs) in quasar spectra with velocity widths larger than 2000 \kms \citep{Anderson87,Weymann91}. Gas fragments shredded by powerful quasar outflows might produce rich multi-component complexes of AALs \citep{Hopkins10, Faucher12b, Chen18, Chen19}. 

These different environments should produce different unique signatures in the absorption-line kinematics, column densities, metal abundances and ionizations. The NALs generally have low metallicities and low ionizations if the gas from the IGM resides at large distances from the quasars. High-speed quasar-driven outflows could exhibit the opposite behavior, with lines showing higher metallicities, higher ionizations, and broader profiles originating from larger column densities of gas in the galactic nuclei. 

%At negative velocity shifts, $z_{abs}< z_{em}$, the absorption lines that form in outflows in quasar environments are mixed with unrelated lines that form in cosmologically intervening gas and galaxies. However, trends could appear in large absorption-line datasets sorted, for example, by velocity shift, line strength, or other quasar properties. 

In this paper, we continue our previous work \citep{Chen20} to investigate the nature and origins of the diverse \civ\ narrow absorption lines (NALs) from high-redshift quasars measured in the Baryon Oscillation Spectroscopic Survey \citep[BOSS,][]{Dawson13, Paris17} of the Sloan Digital Sky Survey III \citep[SDSS-III,][]{Eisenstein11}, which is made possible by a new catalog of quasar absorption lines including \civ\ \lam 1548, 1551 NAL doublets at different velocity shifts developed by York et al. (in prep.) using spectra from BOSS data release 12 (DR12). Our main focus is on the fraction of \civ\ NALs at large velocity shifts, v$\,< -15,000$ \kms\ or v$\, < -40,000$ \kms\ (negative velocities indicate blueshifts relative to the quasar emission-line redshifts), that form in high-speed outflows ejected from the quasars. It is common to assume that all NALs at these extreme velocity shifts form in cosmologically-intervening gas or galaxies unrelated to the quasars. However, detailed studies of individual quasars have shown that some NALs at these large velocity shifts do form in high-speed quasar ejecta \citep{Arav94b, Hamann97d, Misawa07,Simon12}. \citet{Richards99} and \citet{Richards01a} found correlations between the incidence of high-velocity NALs and the quasar radio properties indicating that up to 36\% of these systems form in outflows. 

We reexamine this surprising result by testing the correlations of incidence of high-velocity NALs to the existence of BALs and AALs at lower speeds in the same quasar spectra. We also consider line-locked \civ\ doublets as another indicator of high-velocity NALs formed in outflows. Line-locks are signatures, specifically, of radiative acceleration in outflows because they are believed to result from gas clouds/clumps locked together in velocity by the radiative forces with shadowing effects \citep[see][]{Milne26, Scargle73, Braun89, Ganguly03, Hamann11, Bowler14}. 

In section 2 we describe the quasar samples and the classification of different NAL groups used in this study. Section 3 presents the main results based on both of the correlation analysis and median composite quasar spectra study. Next is Section 4 which discusses the results and the implications for our understanding of high-speed outflows in quasar environments. Finally, we summarize our results from this study in Section 5.

\section{Quasar Samples \& NAL Groups}
We select quasars based on the Quasar Absorption-Line Catalog from York et al. (in prep.), which is created using quasar spectra from BOSS DR12 \citep{Paris17}. The BOSS spectra have a wavelength coverage from $\sim3600$\AA\ through $\sim10,000$~\AA\ at a resolutions of $\sim1300$ in the blue end to $\sim2600$ in the red end \citep{Dawson13}.
We describe our quasar sample selection criteria and \civ\ NALs selection process in \citet{Chen20}. 
We have a full sample of 100,376 quasars, in which 40,696 quasars exhibit a total of 54,154 well-measured \civ\ NALs in their spectra. 
The crux of this study is a statistical comparison between the incidence of high-velocity NALs and other quasar properties. 
%\st{An important part of our analysis is to compare the incidence of a high-velocity NALs to BALs} \citep[described in detail in][]{Chen20} in the same spectra.
%\st{An important part of our analysis is to compare the incidence of AALs and high-velocity NALs to BALs} \citep[described in detail in][]{Chen20} in the same spectra. 

\begin{deluxetable}{cccc}
\tablecaption{Groups of \civ\ NALs. Quantities listed are the group name, velocity shift range (km/s), number of quasars, and number of NALs in each group.\label{tab:definitions}}
\tablehead{
\colhead{Line Group} & \colhead{Velocity shift} & \colhead{\# Quasars} & \colhead{\# NALs}\\
\colhead{}  & \colhead{(\kms)} & \colhead{} & \colhead{}
}  
\startdata
AALs & $-$8000 to 5000 & 22,335 & 25,264 \\
High-v40 & $-$60,000 to $-$40,000 & 8830 & 9432\\
\vspace{1.5mm} High-v15 & $-$60,000 to $-$15,000 & 20,749 & 24,167\\ 
\multirow{ 2}{*}{Line-locks} & $-$60,000 to $-$1000 & \multirow{ 2}{*}{592} & \multirow{ 2}{*}{610 pairs} \\
  & \&\ $450<\Delta v<550$ &  & \\
\enddata
\end{deluxetable}

\Cref{tab:definitions} summarizes the groups we consider, sorted mainly by velocity shift, where negative velocities indicate blueshifts relative to the quasar emission-line redshifts. `AALs' are any NAL within -8000 km/s to 5000 km/s of the emission-line redshift \citep[see more information in][]{Chen20}. We define two groups of high-velocity NALs in the ranges $-$60,000 to $-$40,000 km/s for `high-v40' and $-$60,000 to $-$15,000 km/s for `high-v15'. These groups are dominated by unrelated intervening NALs, but previous studies indicate that a significant fraction of them form in very high-speed quasar-driven outflows \citep{Richards00, Misawa07, Simon12}. We use the high-v40 group specifically to test for relationships and avoid overlap with BAL outflows measured at lower velocity shifts . 

Finally, we define `line-locks' as pairs of \civ\ doublets with velocity separations $\Delta v$ between 450 and 550 km/s corresponding to the \civ\ doublet separation (498 km/s) to an accuracy of 10\%. We also require that all line-locks have velocity shifts v$\, < -1000$ km/s to ensure that they are good outflow candidates and avoid the high-density of AALs near v$\,\sim 0$ that might appear in pairs at the \civ\ doublet separation by random chance \citep[see Figure 2 in][]{Chen20}. Approximately 8\% of quasars with multiple \civ\ NALs at velocity shifts $<-1000$ km/s contains a line-locked NAL pair as defined above.

\begin{figure}
\centering
\includegraphics[width=0.5\textwidth]{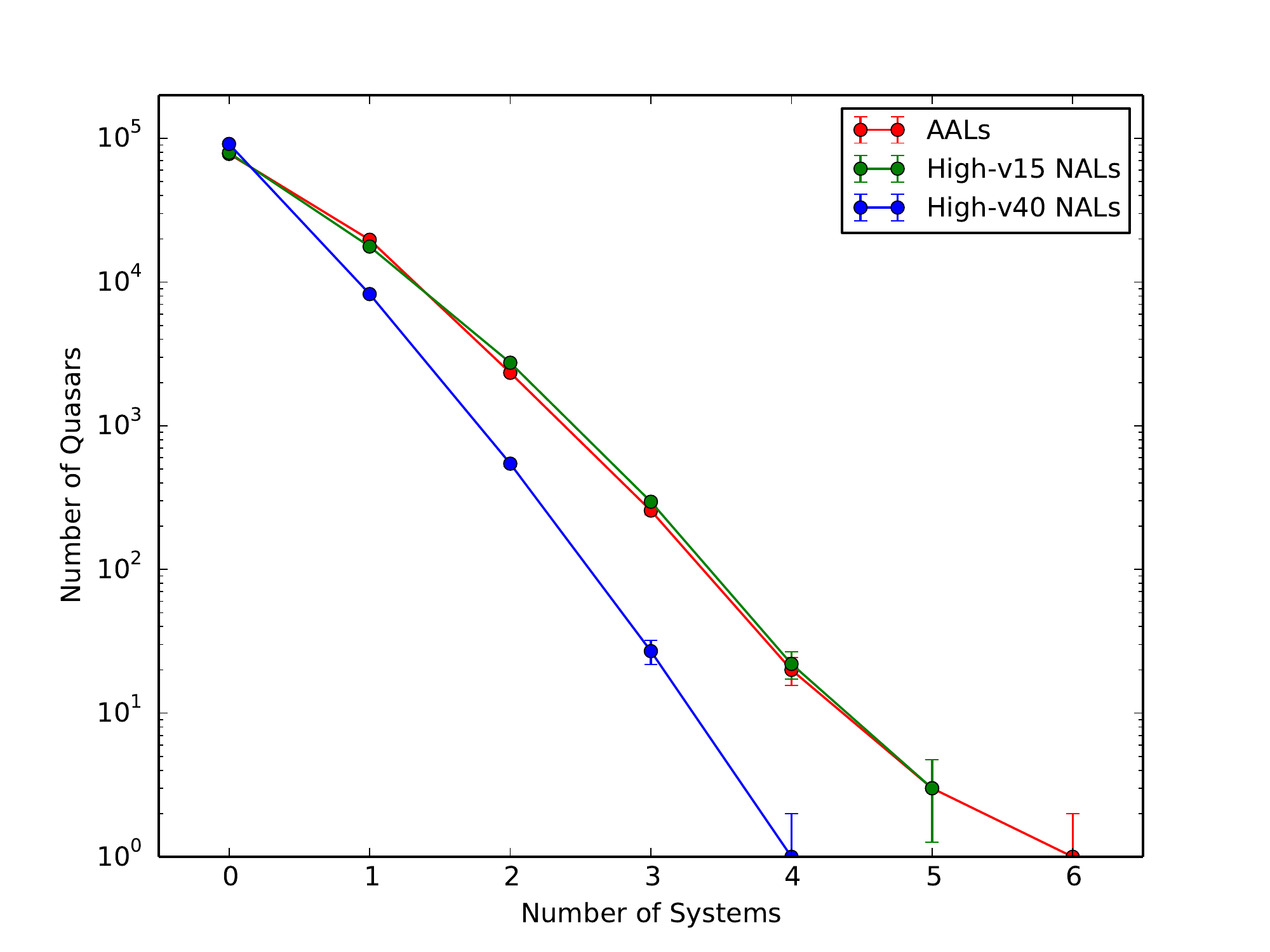}
\caption{The red, green, and blue lines show the numbers of quasars vary with the numbers of AALs, high-v15 NALs, and high-v40 NALs, respectively. The error bars indicate the $1\sigma$ uncertainty from counting statistics.\label{fig:number}}
\end{figure}

\Cref{fig:number} shows the numbers of quasars that have AALs and high-velocity NALs defined above. It is important to keep in mind that these numbers are limited to relatively strong \civ\ NALs with rest equivalent width (REW)~$\geq0.5$ \AA\ due to the relatively low resolution ($R\sim 2000$ corresponding to $\sim$150 \kms) and low signal-to-noise ratios in the BOSS spectra. 

\section{Analysis \& Results \label{sec:correlation}}

\subsection{REW and FWHM Distributions}

\begin{figure*}
\centering
\includegraphics[width=0.8\textwidth]{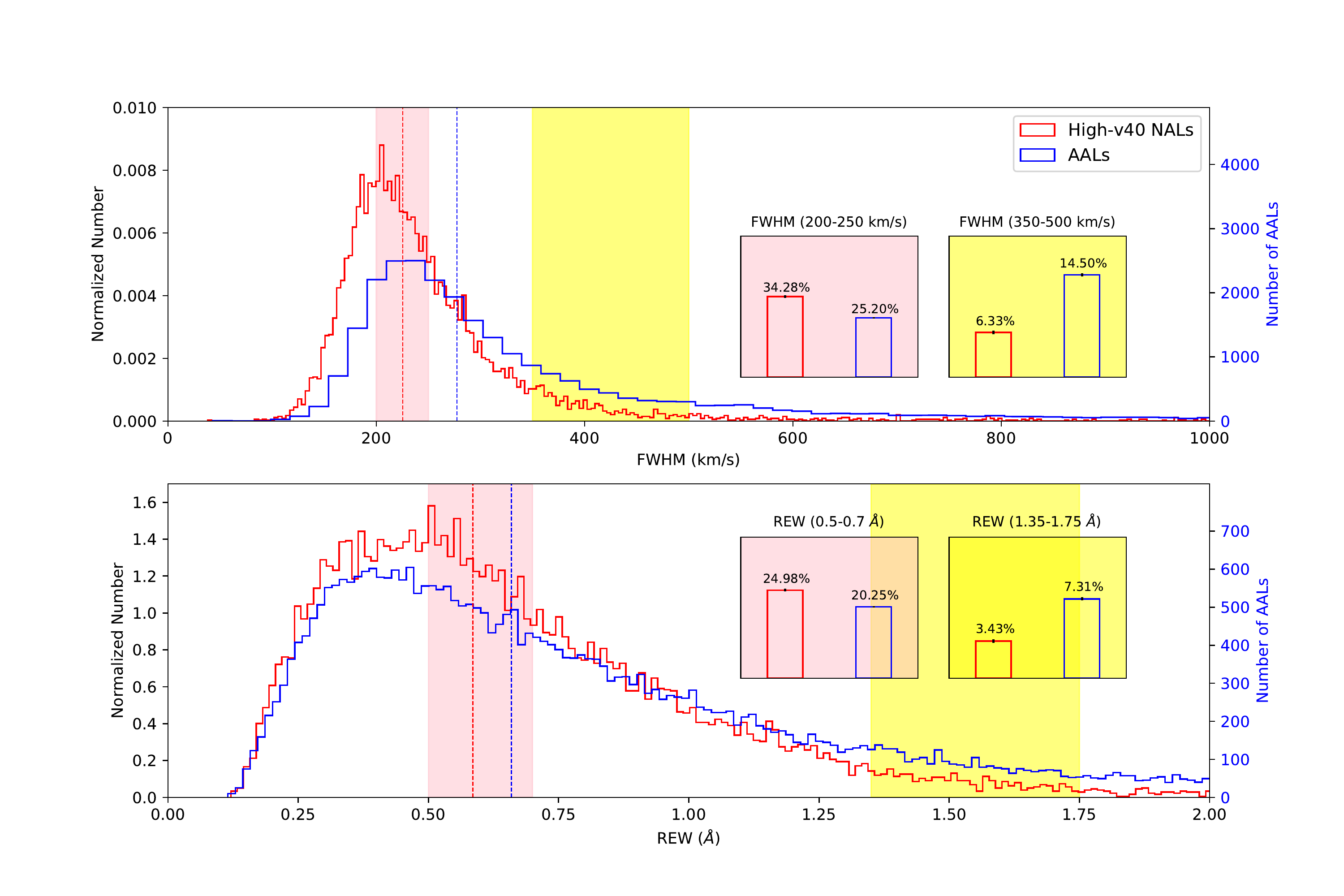}
\caption{Top panel: Normalized FWHM distributions for AALs and high-v40 NALs. The labels of number of AALs in the blue bins are shown on the right axis. Inset: the pink and yellow insets plot the fractions of NALs within the specific color marked areas in the FWHM distribution. The selection of the color marked areas is described in the text. We do not apply the FWHM$\leq500$ \kms\ cutoffs to the FWHM distributions so as to include the full sample of quasars in these plots. The dashed vertical lines show the median values of FWHM distribution, 277 and 225 \kms, in AALs and high-v40 NALs. Bottom panel: Normalized REW distributions for AALs and high-v40 NALs. The labels of number of AALs in the blue bins are shown on the right axis. Inset: the pink and yellow insets plot the fractions of NALs within the specific color marked areas in the REW distribution. We do not apply the REW$\geq0.5$ \AA\ cutoffs so as to include the full sample of quasars in these plots. The dashed vertical lines show the median values of REW distribution, 0.66 and 0.59 \AA, in AALs and high-v40 NALs.\label{fig:rew_fwhm}}
\end{figure*}

\Cref{fig:rew_fwhm} compares normalized FWHM and REW distributions of the \civ\ AALs and high-v40 NALs. The distributions are normalized for easier comparisons. The median values are 0.66, 0.59 \AA\ for the REW distribution of AALs and high-v40 NALs respectively, and 277, 225 \kms\ for the FWHM distribution of AALs and high-v40 NALs respectively. To study the dependence of the NALs occurrence rate on the FWHM and REW distribution, We also compare the fractions of AALs and high-v40 NALs in the lower end (pink marked area) and upper end (yellow marked area) of the FWHM and REW distributions respectively. The reason for selection of the lower end ($200-250$ \kms\ for FWHM distribution and $0.5-0.7$ \AA\ for REW distribution) and upper end ($350-500$ \kms\ for FWHM distribution and $1.35-1.75$ \AA\ for REW distribution) on the plot is described in detail in \citet{Chen20}. As can be seen from the insets, high-velocity NALs, which should be mostly intervening and unrelated to the quasars, tend to be both narrower (from the distribution of FWHM) and weaker (from the distribution of REW) than AALs. 

\subsection{Correlation Analysis}

In the remainder of this section, we study the origins of the extreme high-velocity NALs based on their relationships with the occurrence of AALs and BALs at lower speeds in the same quasar spectra. 

%study the correlations between the incidence rates ofdifferent CivNAL groups to the intrinsic quasar proper-ties including BALs and radio-loudness.

We use the Z-test to determine the statistical significance of correlations between the incidence rates of high-velocity NALs and line-locked NALs to the intrinsic quasar properties including BALs and radio-loudness \citep[see more information in][]{Chen20}. 
For example, when studying the correlation between the incidence rate of high-velocity NALs and the BALs property, we use four numbers as our Z-test inputs, including the number of BALQSOs with and without high-velocity NALs, and non-BALQSOs with and without high-velocity NALs. Similar Z-tests have been done in other correlation analysis.
\Cref{tab:significance} lists the statistical $Z$-value and corresponding significance level for every correlation we compute. We test the correlations between the incidence of high-velocity NALs to BALs, AALs, outflow AALs and radio loudness; and correlations between the incidence of line locked NALs to BALs and radio loudness. We will explain each correlation in detail in the following subsections. The error bars in the plots of this section are $1\sigma$ uncertainty from counting statistics.

\begin{deluxetable}{cccc}
\tablecaption{The Z-test results Z-value (P-values in the parenthesis) of correlation between the incidence rates of high-velocity NALs or line-locked NALs and intrinsic quasar properties, including the presence of BALs, AALs, outflow AALs and radio-loudness as described in the text. Value of $\vert Z\vert \geq 2.58$ indicates a strong correlation at $\geq$99\%\ confidence. For example, the Z-value of 4.94 in the first row suggests BALQSOs are more likely to have high-v40 \civ\ NALs in their spectra than non-BALQSOs.
\label{tab:significance}}
\tablehead{
\colhead{} & \colhead{High-v15} & \colhead{High-v40} & \colhead{Line-locks}
}  
\startdata
 BALs & --- & 4.94 (0) & 17.65$^a$ (0) \\
 AALs & 4.97 (0) & 9.09 (0)  & --- \\
 Outflow AALs & 8.23 (0) & 8.52 (0)  & --- \\
 Radio loudness & 4.43 (0) & 2.85 (0.004) & 1.53 (0.13)\\
\enddata
\tablenotetext{a}{This Z and P-value are measured for line-locked NALs in high-v15 group and BALs with BI$\geq500$ and a velocity cut \texttt{vmin\_civ\_2000}~$> -15,000$ km/s (see Section 3.2.2 for more discussion, and BI and \texttt{vmin\_civ\_2000} cuts are described in detail in \citet{Chen20}).}
\end{deluxetable}

\subsubsection{High-velocity NALs}

\begin{figure}
\centering
\includegraphics[width=0.5\textwidth]{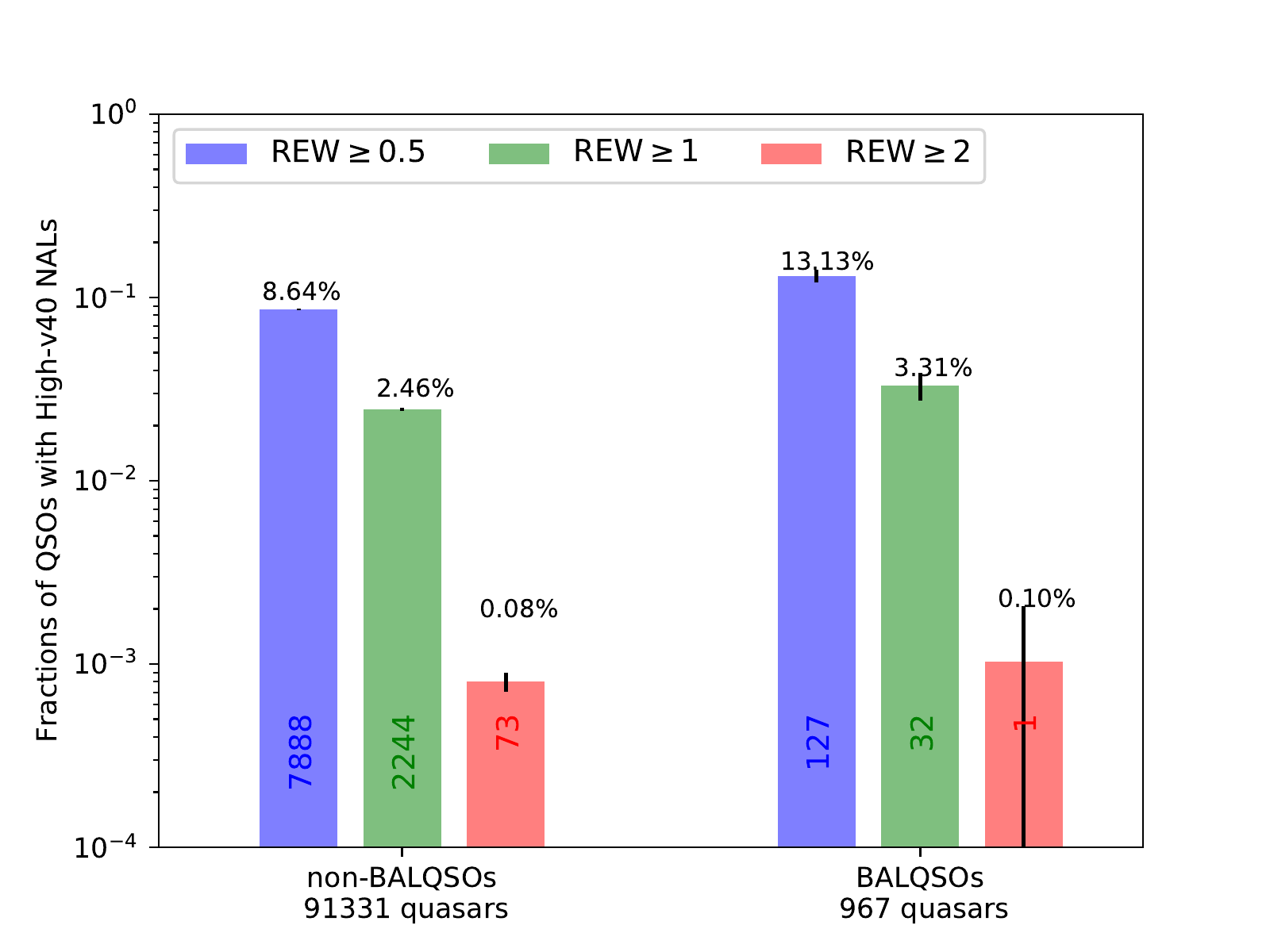}
\caption{Fractions of quasars with high-v40 NALs in both non-BALQSOs and BALQSOs in different REW thresholds(REW$\geq$0.5, REW$\geq$1, and REW$\geq$2). The number of quasars with high-v40 NALs is labeled on each bar, and the fraction of quasars with high-v40 NALs is labeled on the top of each bar. The total numbers of non-BALQSOs and BALQSOs included in these plots (91,331 and 967, respectively) are recorded below the panels. The error bars are $1\sigma$ uncertainty from counting statistics.\label{fig:BAL_inter}}
\end{figure}

We examine NALs at large velocity shifts to determine if a fraction of them form in high-speed quasars instead of cosmologically-intervening gas or galaxies. Correlations with other quasar properties would provide direct evidence for an outflow origin \citep[e.g.,][and Section 1]{Richards99, Richards01a, Stone19}. \Cref{fig:BAL_inter} shows the fractions of quasars with high-velocity \civ\ NALs in the high-v40 group for BALQSOs and non-BALQSOs. There is a significant relationship of high-v40 NALs to the appearance of BALs in the same spectrum. The probability of this correlation occurring by chance is $P\approx 0\%$ (\Cref{tab:significance}). There is no significant dependence on the NAL REWs. It is necessary to note that BALQSOs tend to have higher signal-to-noise spectra than the non-BALQSOs (the median value of SNR$_{1700}$ for non-BALQSOs is $\sim$7.5, and the median value of SNR$_{1700}$ for BALQSOs is $\sim$10.0), which could affect the statistics.

\begin{figure*}
\centering
\includegraphics[width=1.0\textwidth]{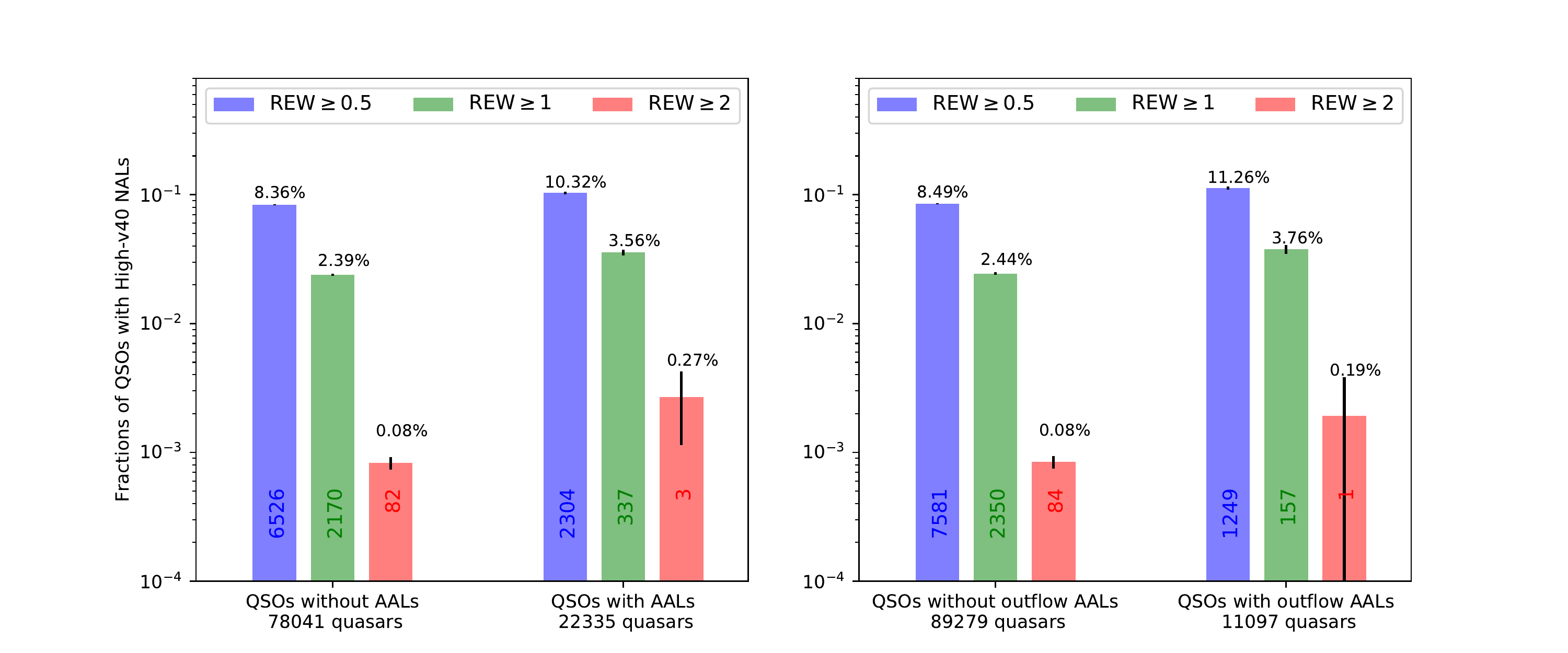}
\caption{Fractions of quasars with high-v40 NALs in quasars with AALs (left panel) and outflow AALs (right panel). The number of quasars with high-v40 NALs is labeled on each bar. See \Cref{fig:BAL_inter} for additional notes.\label{fig:AAL_inter}}
\end{figure*}

\Cref{fig:AAL_inter} plots the fractions of quasars with one or more high-v40 NALs for quasars with and without AALs (left panel) and with and without outflow AALs (right panel). There are strong correlations between high-v40 NALs and both AALs and outflow AALs, with the probability of these correlations occurring by chance $P\approx 0\%$ (\Cref{tab:significance}). The results are similar for both AALs and outflow AALs, again with no significant dependence on the high-v40 NAL REWs. We also test the correlations for high-v15 NALs to AALs and outflow AALs, which could improve the statistics. It gives us similar results as high-v40 that there are also strong correlations between high-v15 NALs and both AALs and outflow AALs, with the probability of these correlations occurring by chance $P\approx 0\%$ (\Cref{tab:significance}). 

If we assume that all high-v40 NALs related to outflow AALs form in high-speed quasar-driven outflows in the quasar environments, then the fraction of outflow systems in the high-v40 group (with any REW $\geq$ 0.5 \AA ) is $\sim$25\%, and the fraction that is intervening and unrelated to the quasars is $\sim$75\%. And If we apply this to high-v15 NALs, then the fraction of outflow systems in the high-v15 group (with any REW $\geq$ 0.5 \AA ) is $\sim14$\%, and the fraction that is intervening and unrelated to the quasars is $\sim86$\%. If we assume that all high-v40 NALs related to BALs form in quasar-driven outflows, then the fraction of outflow systems in the high-v40 group (with any REW $\geq$ 0.5 \AA ) is $\sim$34\%, and the fraction that is intervening and unrelated to the quasars is $\sim$66\%. All the results indicate that a significant fraction of these systems is ejected from the quasars.

\begin{figure*}
\centering
\includegraphics[width=1.0\textwidth]{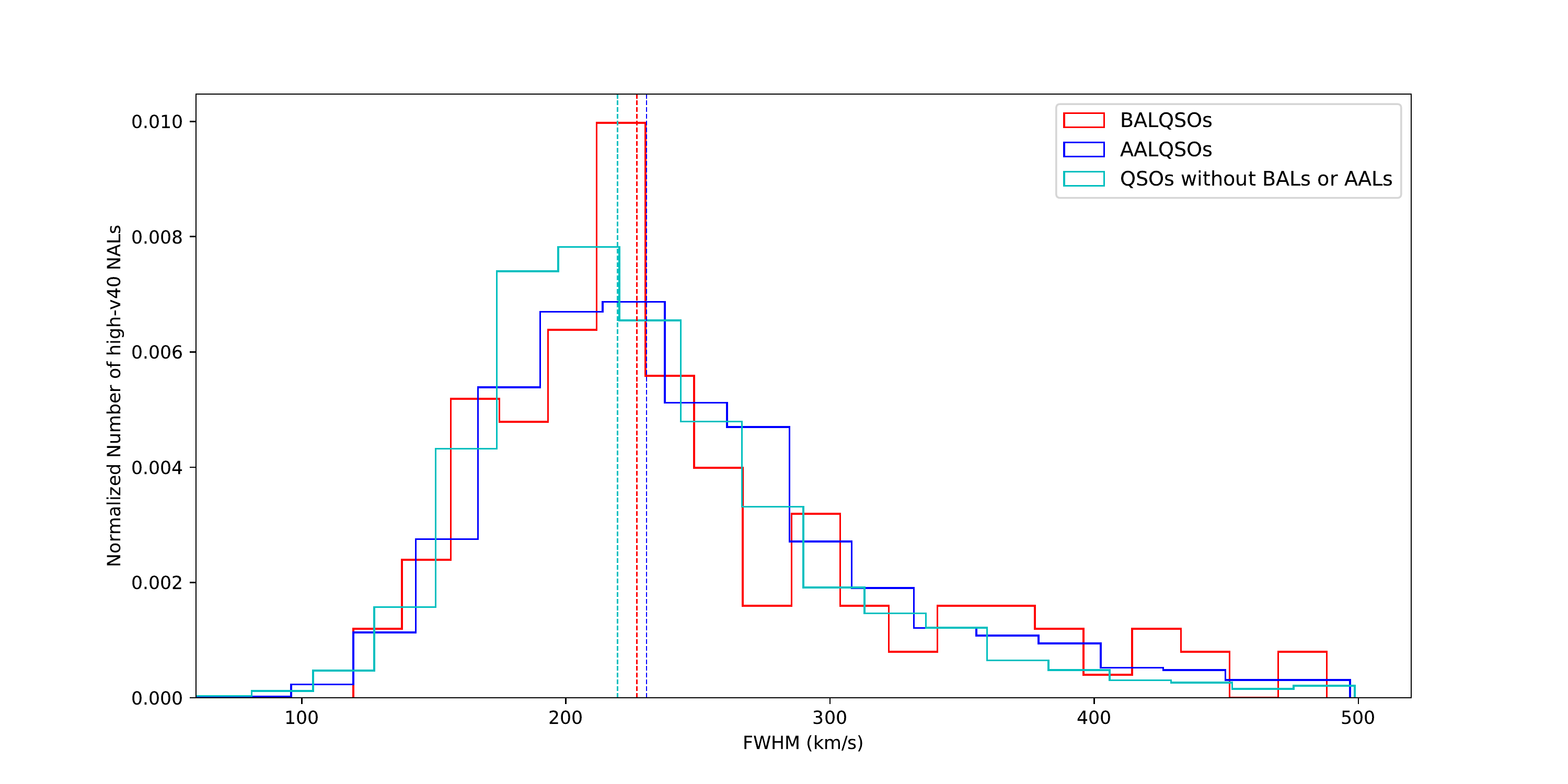}
\caption{Normalized FWHM distributions for high-v40 NALs in BALQSOs, AALQSOs, and quasars without BALs or AALs, respectively. The dashed vertical lines show the median values of FWHM (227, 231, and 220 \kms) in the same color code.\label{fig:fwhm_highv}}
\end{figure*}

\Cref{fig:fwhm_highv} plots FWHM distributions for high-v40 NALs in BALQSOs, AALQSOs, and quasars without BALs or AALs, respectively. The distributions are normalized to the same scale for easier comparisons. The dashed vertical lines in \Cref{fig:fwhm_highv} show the median value for each distribution. One obvious result from this plot is that high-velocity NALs in BALQSOs and AALQSOs tend to have larger FWHMs.

\subsubsection{Line-Locked NALs}

\begin{figure*}
\centering
\includegraphics[width=1.0\textwidth]{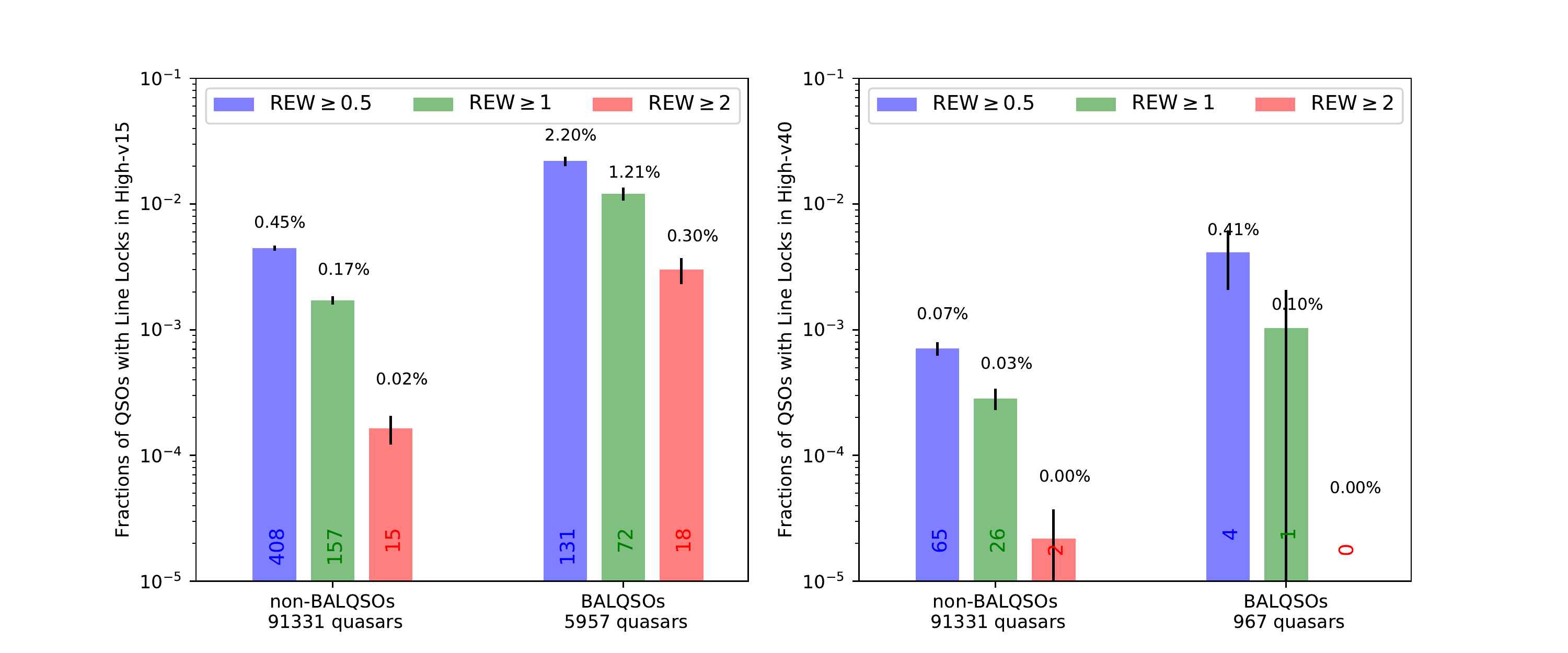}
\caption{Fractions of quasars with high-v15 line-locked NALs (left panel), and high-v40 line-locked NALs (right panel) in both non-BALQSOs and BALQSOs. For the BALQSOs in left panel, we define BAL differently by requiring \texttt{vmin\_civ\_2000}~$> -15,000$ km/s to avoid line-locked NALs sitting inside the BAL troughs. The number of quasars with line locked NALs is labeled on each bar. See \Cref{fig:BAL_inter} for additional notes.\label{fig:linelock}}
\end{figure*}

Line-locked NALs are an interesting separate class of NALs because, at the velocity shifts we consider, v$\,<-1000$ \kms , they should form in quasar outflows driven by radiative forces (Section 1). \Cref{fig:linelock} shows the fractions of quasars with a line-locked NAL pair in the high-v15 group (left panel) and a line-lock pair in the high-v40 group (right panel) for BALQSOs versus non-BALQSOs. We only study the line locks in high-velocity NALs to avoid line-locked NALs sitting inside the BAL troughs and the high-density of outflow AALs (with velocities at $-8000$ to $-1000$ \kms) that might appear in pairs at the \civ\ doublet separation by random chance. We also define BALs in the left panel differently by requiring BI$\geq500$ and \texttt{vmin\_civ\_2000}~$> -15,000$ km/s to avoid line-locked NALs sitting inside the BAL troughs (see \citet{Chen20} for descriptions of BI and \texttt{vmin\_civ\_2000} cuts). There is a strong relationship between all line-locked NALs and BALs, with the probability of this correlation occurring by chance $P=0\%$, listed in \Cref{tab:significance}. BALQSOs are more likely to have line-locked NALs than non-BALQSOs. These correlations are stronger at larger REW thresholds.

\subsubsection{Comparisons to Radio Loudness}

Radio-loudness in quasars is an indication of relativistic jets, which produce the radio flux via synchrotron emissions. These radio jets might sweep up cooler gas in the quasar host galaxies to produce outflow NALs when viewed along particular lines of sight. We compare the fractions of quasars with high-velocity NALs and line locks versus radio loudness. The results are listed in \Cref{tab:significance}. There is a a strong correlation between radio loudness and high-v40 NALs, with $P=0.4\%$, and a stronger correlation between radio loudness and high-v15 NALs, with $P=0\%$. There is a weak correlation between radio loudness and line locks, with $P=13\%$. We will discuss these results in Section 4 below.

\subsection{Composite Spectra}

\begin{figure}
\centering
\includegraphics[width=0.45\textwidth]{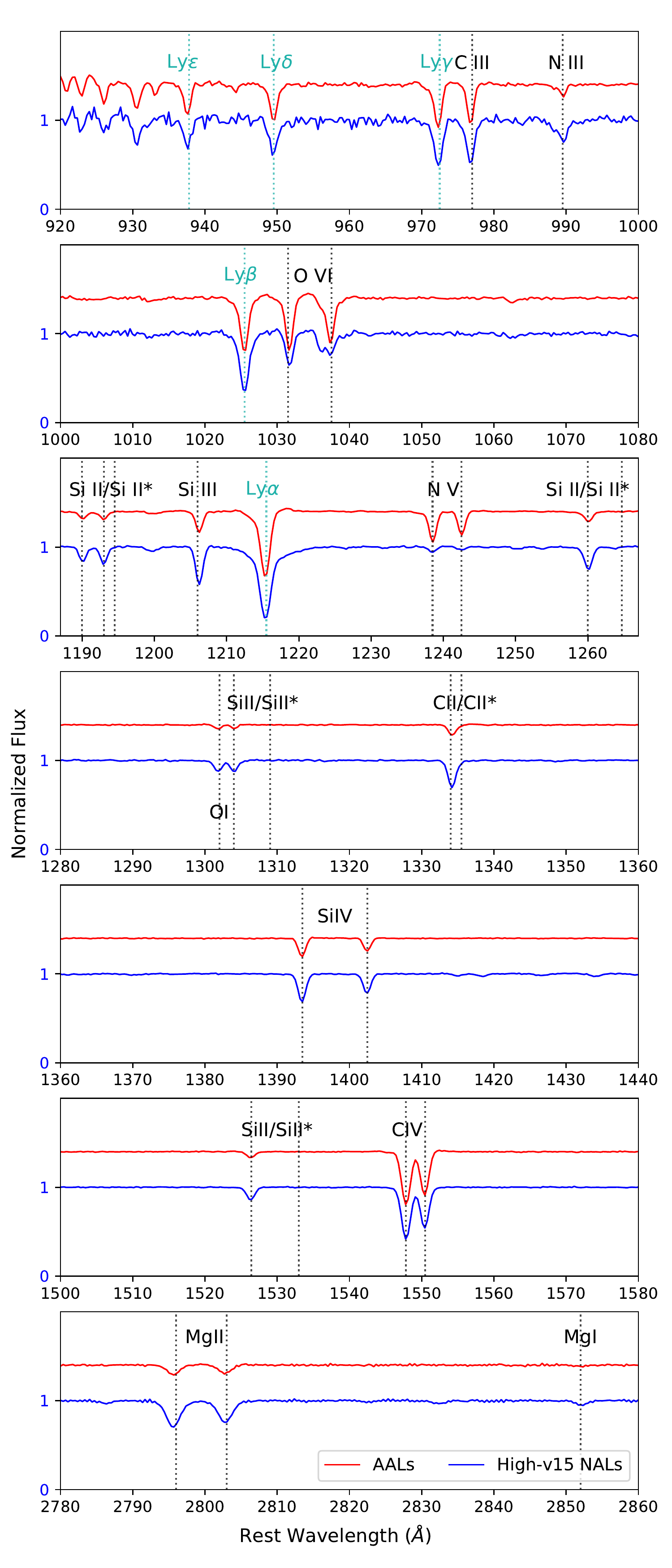}
\caption{Normalized median composite spectra of non-BALQSOs in the absorber frame sorted by AALs and high-v15 NALs. The red and blue lines show the composite spectra of AALs and high-v15 NALs respectively. Offsets are added to the spectra for displaying purpose.\label{fig:composite_norm}}
\end{figure}

We create median composite spectra in the absorber frame for all non-BAL quasars sorted by AALs and high-v15 NALs. Portions of the resulting composites are shown in \Cref{fig:composite_norm} after normalizing to unity in the continuum. The most obvious differences between groups in this figure are that the high-velocity NALs (high-v15) have stronger absorption in low-ionization lines such as \siii\ \lam 1190, 1260, 1527, \cii\ \lam 1335, \mgii\ \lam 2796, 2804, and \mgi\ \lam 2853, weaker absorption in high-ionization lines such as \ovi\ \lam 1032, 1038, and much weaker absorption in \nv\ \lam 1239, 1243 compared to the AALs. This is consistent with lower ionizations in the high-velocity NALs dominated by intervening systems far from the quasars. We also find that high-velocity NALs show lower \nv/\civ\ and \nv/\ovi\ absorption ratios than the AALs, but similar or higher \niii/\ciii\ absorption ratios. We will discuss these results further in Section 4 below.

\begin{figure}
\centering
\includegraphics[width=0.45\textwidth]{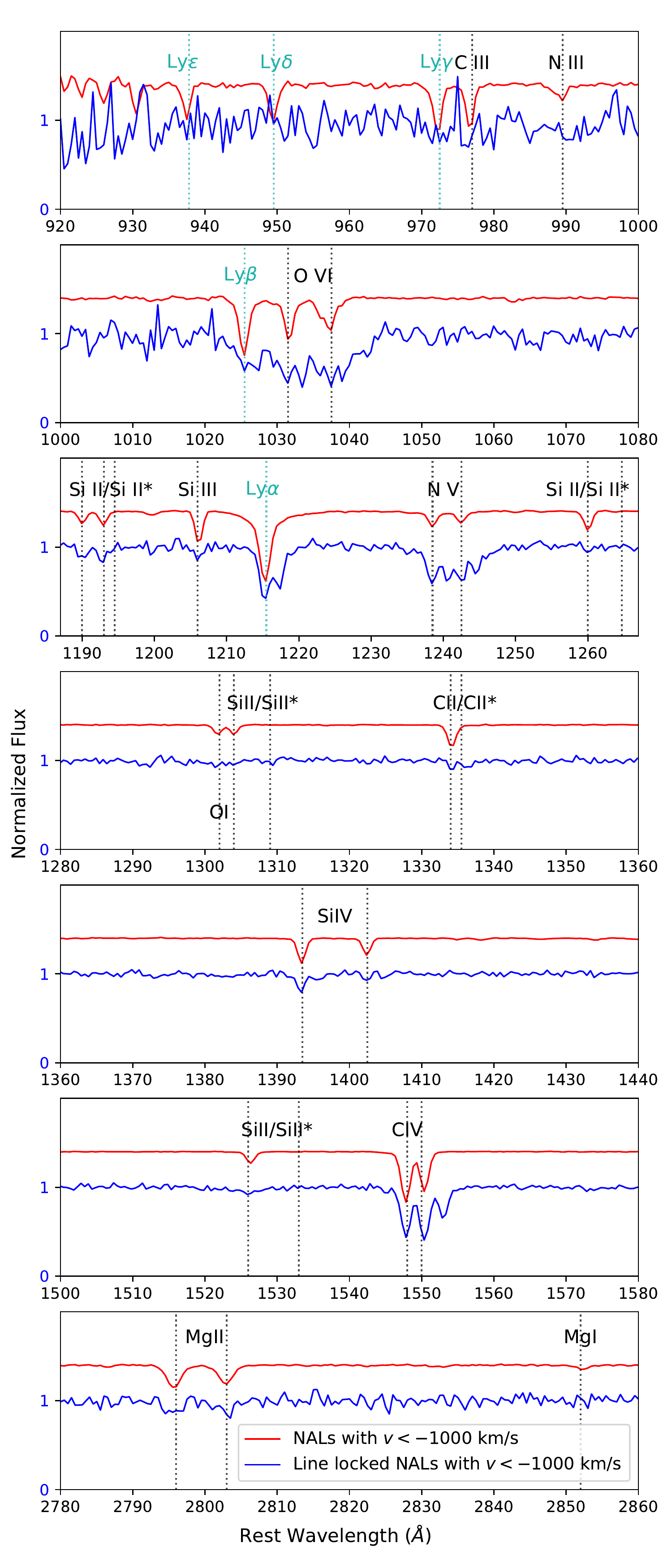}
\caption{Normalized median composite spectra of non-BALQSOs in the absorber frame of \civ\ NALs. The red and blue lines show the composite spectra of overall NALs and line-locked NALs, respectively, both at $v<-1000$ km/s. See additional notes in \Cref{fig:composite_norm}.\label{fig:lock_composite_norm}}
\end{figure}

We also create median composite spectra for all non-BALQSOs with overall and with line-locked NALs at velocity shifts v$< -1000$ km/s. \Cref{fig:lock_composite_norm} compares these two matched composites. The line-lock NAL shows a trio of \civ\ lines that are the line-locked pair, plus the expected splitting in other detected lines at the \civ\ doublet separation. The NAL systems in both composites show a wide range of ionizations from \mgii\ to \nv\ and \ovi. However, the line-locked NALs have generally higher ionizations than NALs overall based on weaker absorption in low-ionization lines like \lya, \mgii, \cii, and \siii\ and stronger absorption in high-ionization lines such as \nv\ and \ovi . 

\section{Discussion}

In Section 3, we described comparisons between high-velocity \civ\ NALs, line-locked NALs, and their relationships to others quasar properties, namely BAL outflows and radio-loudness. We find numerous strong correlations as well as differences in the line ratios (ionizations) that provide new constraints on the physical nature of outflows in quasar environments. Here we provide a brief discussion.

A surprising result is that the incidence of high-velocity NALs at shifts v$\,< -40,000$ \kms\ (in our high-v40 group) is significantly larger in quasars with measured BAL outflows (Section 3.2.1). This relationship indicates that $\sim$25\% of all high-velocity NALs in our study form in very high-speed quasar-driven outflows and, conversely, only $\sim$75\% form in unrelated intervening gas or galaxies. We also find that high-velocity NALs in BALQSOs and AALQSOs tend to have larger FWHMs (Section 3.2.1), which is consistent with the previous result that a significant fraction of the high-velocity systems is ejected from the quasars instead of forming in unrelated intervening gas or galaxies \citep{Misawa07, Simon12}. Besides, we find a significant correlation between high-velocity NALs and radio loudness (Section 3.2.3). These results confirm previous work by \citet{Richards99} and \citet{Richards01a} who used correlations with quasar radio properties to infer that up to $\sim$36\% of high-velocity \civ\ NALs form in outflows. They stressed the need for truly homogeneous and unbiased surveys to test their results. Our study provides this confirmation in the BOSS dataset that is unbiased for NAL properties. Our study also confirms the most recent work by \citet{Stone19}, who estimated that $\sim$30\%\ high-velocity NALs are intrinsic. We need to note that high-velocity NALs are correlated with both BAL outflows and radio-loudness, but BALQSOs are generally radio quiet \citep[discussed in][]{Chen20}.

The large fractions of high-velocity NALs formed in quasar outflows could be a concern for studies of the IGM and intervening galaxies that use metal absorption lines in quasar spectra. ``Associated'' absorption lines were originally defined in studies like this as a separate class at $z_{abs}\approx z_{em}$ that is likely to be physically related to the quasars versus other NALs at high velocities that are generally assumed to form in cosmologically intervening material. The definition of AALs has expended in recent years to include higher-velocity systems at shifts up to v$\,\sim -8000$ \kms\ or v$\,\sim -10,000$ \kms\ based on statistical studies of large samples (see Figure 2 in \citealt{Chen20}, also \citealt{Nestor08}, \citealt{Wild08}). The results presented here indicate that NALs formed in quasar outflows appear significantly at all velocity shifts in quasar spectra from v$\,\sim 0$ to v$\,\sim 0.2c$. The relationship to radio loudness might imply that some of these high-speed outflow NALs form in material swept up by the radio jets. Our composite spectra show that the high-velocity NALs have typically lower ionizations than all types of AALs (Section 3.3). We attribute this to lower ionizations in true intervening absorbers. These absorbers still dominate the high-velocity NAL samples in our study and they should naturally have lower ionizations than AALs because they are photoionized by the diffuse inter-galactic background radiation instead of the harsh radiation field of the quasars \citep[see also][]{Perrotta18}. From previous studies, the \nv/\civ\ absorption ratio is sensitive to the metallicity \citep[e.g.,][]{Hamann99}. Our composite spectra show lower \nv/\civ\ and \nv/\ovi\ absorption ratios in high-velocity NALs than all of the AAL groups, which could be caused by metallicities (enhanced N/C and N/O) near quasars, but similar \niii/\ciii\ absorption ratios suggest it is caused by ionization effects. The ratio of \nv/\civ\ absorption could naturally decrease farther from quasars (intervening systems) because the gas there is less ionized (not exposed to the harsh quasar radiation field), but then it is surprising to strong \ovi\ absorption in high-velocity NALs. Previous studies use a hot gas component with $T > $300, 000 K that can support \ovi\ and some higher ions but not \nv\ (nor \civ\ nor lower ions) to explain this phenomenon. So \civ\ absorption is produced in intervening gas by photoionization by the UV background, and \ovi\ absorption is produced in this hot thermal component. \nv\ is weak in absorption because it falls in an ionization gap between these two components. 

Another indicator of NALs formed in quasar outflows is pairs of NALs locked together at the \civ\ doublet separation. The lines become locked at the doublet separation due to shadowing effects in radiatively-driven outflows (see refs in Section 1). Our analysis shows that roughly 8\%\ of quasars in our sample with multiple \civ\ NALs at $v<-1000$ km/s have at least one line-lock pair, and the percentage increases to $\sim15\%$ at velocities within -10,000 km/s, which suggest that line locks are common in quasar outflows. 

The composite spectra of line locks (\Cref{fig:lock_composite_norm}) indicate that the line-locked NALs have generally higher ionizations than NAL systems at the same velocity shifts. We also find that the incidence of line-locked NALs correlates with both BAL outflows and radio loudness. All of these results are consistent with the majority of observed line-locked NALs forming in quasar outflows accelerated by the radiation field of the quasars.

\section{Summary}

We use SDSS-BOSS DR12 database to investigate the nature and origins of high-velocity \civ\ NAL outflows (with speeds $\sim 0.1 -0.2c$) by studying their relationships to other quasar properties such as the incidence of AALs, BAL outflows, and radio-loudness. We also study the properties of line-locked NALs in the same quasar sample. Our analysis yields the following results:

1) High-velocity NALs are strongly correlated with both AALs and BALs, indicating a significant fraction ($\sim25\%$) of the extremely high-velocity systems is ejected from the quasars, and only $\sim75\%$ form in unrelated intervening gas or galaxies (Section 3.2.1 and 4).

2) High-velocity NALs are also strongly correlated with radio loudness. This might imply that some of these high-speed outflow NALs form in material swept up by the radio jets (Section 3.2.3 and 4).

3) High-velocity NALs have typically lower ionizations than all types of AALs. We attribute this to lower ionizations in true intervening absorbers. These absorbers still dominate the high-velocity NAL samples in our study, which is consistent with the above result 1. The small \nv/\civ\ and \nv/\ovi\ absorption ratios in high-velocity NALs are caused by ionizations rather than metallicities (Section 3.3 and 4). 

4) Line-locked NALs are correlated with both BAL outflows and radio loudness (Section 3.2.2). And they have weak low-ionization absorptions and strong high-ionization absorptions, suggesting that line-locked NALs are highly ionized (Section 3.3). The facts that they are highly ionized and correlated to both BAL outflows and radio-loud quasars imply physical line-locking due to radiative forces is real (Section 4).

\section*{acknowledgments}
We are grateful to Britt Lundgren, Don York, Yusra Alsayya for providing us with their unpublished quasar catalog, which is the foundation of this work. CC and BM acknowledge financial support from the Fundamental Research Funds for the Central Universities (Sun Yat-sen University) and NFSC grant 12073092. FH acknowledge the support of funds from University of California, Riverside, USA and by grant AST-1009628 from the USA National Science Foundation.

\bibliography{reference}
\bibliographystyle{aasjournal}

\end{document}